# A magnetically-induced Coulomb gap in graphene due to electron-electron interactions

Evgenii E. Vdovin[1,2], Mark T. Greenaway[3 ✉], Yurii N. Khanin[2], Sergey V. Morozov[2], Oleg Makarovsky[1], Amalia Patanè[1], Artem Mishchenko[4,5], Sergey Slizovskiy[4,5], Vladimir I. Fal'ko[4,5,6], Andre K. Geim[4,5], Kostya S. Novoselov[7] & Laurence Eaves[1]

Insights into the fundamental properties of graphene's Dirac-Weyl fermions have emerged from studies of electron tunnelling transistors in which an atomically thin layer of hexagonal boron nitride (hBN) is sandwiched between two layers of high purity graphene. Here, we show that when a single defect is present within the hBN tunnel barrier, it can inject electrons into the graphene layers and its sharply defined energy level acts as a high resolution spectroscopic probe of electron-electron interactions in graphene. We report a magnetic field dependent suppression of the tunnel current flowing through a single defect below temperatures of ~2 K. This is attributed to the formation of a magnetically-induced Coulomb gap in the spectral density of electrons tunnelling into graphene due to electron-electron interactions.

[1] School of Physics and Astronomy, University of Nottingham, Nottingham NG7 2RD, UK. [2] Institute of Microelectronics Technology RAS, Chernogolovka 142432, Russia. [3] Department of Physics, Loughborough University, Loughborough LE11 3TU, UK. [4] Department of Physics and Astronomy, University of Manchester, Manchester M13 9PL, UK. [5] National Graphene Institute, University of Manchester, Manchester M13 9PL, UK. [6] Henry Royce Institute for Advanced Materials, University of Manchester, Manchester M13 9PL, UK. [7] Institute for Functional Intelligent Materials, National University of Singapore, 4 Science Drive 2, Singapore 117544, Singapore. ✉email: m.t.greenaway@lboro.ac.uk

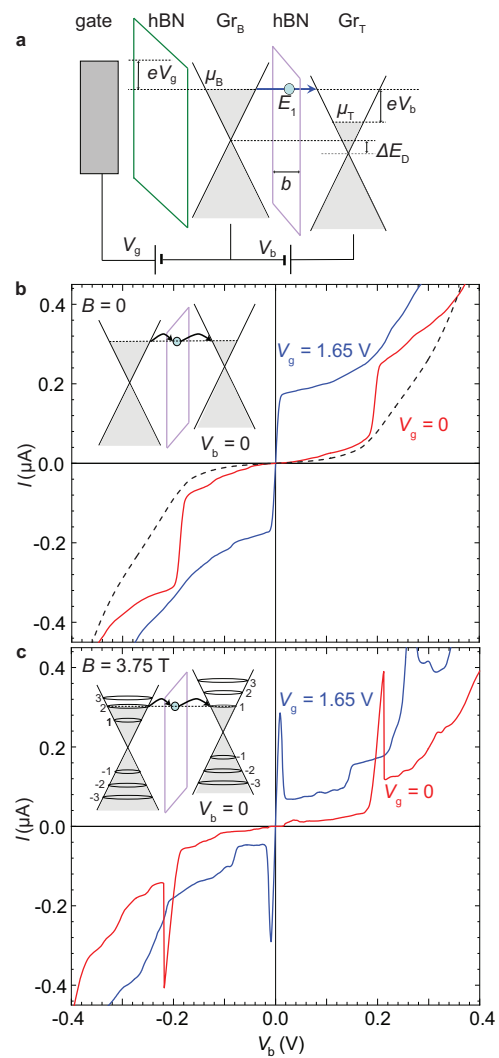

Electron-electron (e-e) interactions have a significant effect on the resistivity, $\rho$, of planar graphene transistors. For example, in graphene-on-hBN superlattices, Umklapp scattering by atoms in the crystal lattice can provide a pair of Dirac fermions with a large momentum "kick" which decreases the conductivity[1]. In contrast, the ballistic flow of electrons through a narrow graphene constriction leads to a significant increase of the conductance due to the viscosity induced by e-e scattering[2]. In magic angle twisted bilayer graphene, strong electron-electron interactions result in the emergence of superconductivity[3] and correlated insulating phases[4]. In double-layer graphene devices with an ultra-thin hBN spacer layer, inter-layer e-e interactions produce a strong Coulomb drag on the electrical current[5] and the formation of exciton condensates[6].

Here we investigate e-e interactions in tunnel transistors in which the current flows by electron tunnelling through a few monolayers of hBN sandwiched between two graphene layers[7]. When the lattices of the graphene layers are aligned so that their relative angular misorientation (twist angle) is small, ≲3°, electrons can tunnel coherently between the Dirac cones of the two layers with the conservation of momentum[8–15]. When the twist angle is large, an electron must scatter in order to tunnel between the misaligned Dirac cones; in this case, tunnelling can occur by phonon-assisted electron tunnelling[16–18] or by hopping via defects or impurities within the tunnel barrier[18–26]. The large twist angle of ≈30° between the two graphene layers in our devices ensures that direct band-to-band tunnelling with conservation of in-plane momentum is suppressed.

A high quality hBN tunnel barrier typically contains within its band gap only a small number of strongly localised defect states, each with a sharply-defined energy level. Such a quantised state can provide a channel through which electrons can tunnel resonantly from one graphene layer to the other[18–26]. We have selected a device with a single active localised state within the tunnel barrier. This state enables electrons to be injected into the Fermi liquid of a graphene layer. Our gated devices allow us to tune the energy of the state and the chemical potential of the graphene layers, thus providing a high resolution spectroscopic probe to investigate the effect of e-e interactions on the tunnel current. Here, we show that at low bias voltages and in a quantising magnetic field, these interactions give rise to a marked reduction in the inter-Landau level tunnel current which emerges at temperatures below ~2 K due to the formation of a 10 meV-wide Coulomb gap in the spectral density of graphene's Dirac fermions. The formation of a Coulomb gap has been studied previously for electrons in GaAs quantum well nanostructures[27–38]. Our analysis based on that earlier work can account for the magnetic field dependence of the Coulomb gap generated in graphene at low temperatures.

### Results and discussion

**Device structure.** Figure 1a shows a schematic diagram of our multi-layer van der Waals heterostructure. It is one of a series of similar devices fabricated by mechanical exfoliation and transfer of the component graphene and hBN flakes onto a 0.5 mm-thick quartz substrate. A 30 nm-thick graphite layer mounted on the substrate acts as the transistor's gate electrode. This layer is capped by a 27 ± 2 nm thick insulating layer of hBN. The tunnel device on the surface of the insulating layer consists of a monolayer of graphene on each side of a 4 monolayer hBN barrier with thickness $b = 1.5$ nm. Electrical contacts to the graphite and graphene layers were made using electron beam lithography and deposition of Cr/Au layers. see methods section. The tunnel current flows through an area of ≈50 μm² formed by the overlap of the top ($Gr_T$) and bottom ($Gr_B$) graphene electrodes.

**Fig. 1 Tunnel current through a localised state in a graphene-hBN tunnel transistor. a** Schematic band diagram of the device showing the energy level $E_1$ of the localised state in the barrier. Here, $V_g$ and $V_b$ are the applied gate and bias voltages, $\mu_B$ and $\mu_T$ are the chemical potentials on the bottom and top graphene layers measured with respect to the Dirac point on each layer, $b$ is the tunnel barrier thickness and $E_D$ is the energy difference between the Dirac points of each layer. **b** Measured steps in current-voltage, $I(V_b)$, curves at $V_g = 0$ (red) and $V_g = 1.65$ V at a temperature $T = 1.75$ K and in zero magnetic field, $B = 0$, due to tunnelling through a single defect state. To compare the relative magnitude of the tunnelling current through one defect, the dashed black curve shows $I(V_b)$ for a similar defect-free tunnel device. **c** $I(V_b)$ curves at finite magnetic field, $B = 3.75$ T, $T = 1.75$ K and $V_g = 0$ (red) and $V_g = 1.65$ V (blue): the inset shows schematically the current flow when $V_b = 0$ and $V_g > 0$.

This arrangement enables us to identify and study the resonant tunnel current flowing through the energy level of the single localised state within the hBN barrier.

The gate, $V_g$, and bias, $V_b$, voltages control the chemical potentials, $\mu_B$ and $\mu_T$, in the bottom ($Gr_B$) and top ($Gr_T$) graphene layers, and the electric field, $F_b$, in the barrier, see Fig. 1a. We define $\mu_B$ and $\mu_T$ relative to the Dirac points in the bottom and top graphene layers respectively. The applied voltages give rise to an energy difference, $\Delta E_D = ebF_b$, between the Dirac points of the two graphene layers and to a potential difference between them, so that $eV_b = \mu_B - \mu_T - \Delta E_D$. Under these conditions, electrons tunnel through the defect from the filled states in one graphene electrode into the empty states of the other, giving rise to the measured tunnel current[7,17]. Details of the electrostatic model used to calculate $\mu_B$ and $\mu_T$ as functions of $V_g$ and $V_b$ are presented in Supplementary Note 1.

**Current-voltage characteristics when $B = 0$.** Figure 1b shows the current-voltage characteristics of the device at low temperatures (1.75 K) and at two different gate voltages with $B = 0$. The steps on the red and blue $I(V_b)$ curves arise from electrons tunnelling through a single localised state in the barrier with energy, $E_1$, measured relative to the Dirac point of the bottom graphene electrode $Gr_B$. The energy $E_1$ depends on the electric field in the barrier, $F_b$, so that $E_1 = eF_b z_1 + E_1^0$ where $z_1$ is the position of the localised state measured from the edge of the barrier and $E_1^0$ is the energy of the state measured relative to the Dirac point of $Gr_B$ when $F_b = 0$, see Fig. 1a. Using our electrostatic model to analyse the features in the $I(V_b)$ curves, we find that $z_1 \approx b/2 = 0.75 \pm 0.05$ nm and $E_1^0 = 90 \pm 10$ meV. The red curve plots $I(V_b)$ when $V_g = 0$. As $V_b$ is increased, $E_1$ decreases and $\mu_B$ increases. When $V_b \approx 0.2$ V, $\mu_B$ aligns energetically with $E_1$ so that a resonant tunnelling channel opens through this state (blue arrow in Fig. 1a). This generates a strong additional contribution to the tunnel current in the form of a large step-like feature on the $I(V_b)$ curve[18–20,39]. The symmetric positions and heights of the steps at positive and negative $V_b$ in Fig. 1b, are consistent with the defect's location being close to the central plane of the tunnel barrier. At $V_g = 1.65$ V the step in $V_b$ is shifted from $\approx 0.2$ V down to a few mV. The gate enables us to modify the electrostatics of the device and to align $E_1$, $\mu_B$ and $\mu_T$ at zero bias, see inset of Fig. 1b[20]. The dashed black curve in Fig. 1b shows $I(V_b)$ at $V_g = 0$ for a similar device. In that device, no step-like features are present, which we attribute to the absence of current-carrying defect states within its tunnel barrier.

**Effect of Landau quantisation on the current-voltage characteristics.** We now examine the effect of a magnetic field, $B$, applied perpendicular to the graphene layers. It transforms the energy spectrum of the carriers into a set of discrete energy levels with energies $E_{N_{B,T}} = \sqrt{2N_{B,T}}\hbar v_F/l_B$, where $N_{B,T}$ is the Landau level (LL) index of the bottom and top layers, $v_F$ is the Fermi velocity of graphene and $l_B = \sqrt{\hbar/eB}$ is the magnetic length (see inset of Fig. 1c).

The red curve in Fig. 1c plots $I(V_b)$ when $T = 1.75$ K, $V_g = 0$ and $B = 3.75$ T. The Landau level quantisation transforms the step increase in $I(V_b)$ (Fig. 1b) at $V_b \approx 0.2$ V and $B = 0$ into a resonant peak. Similarly, when $V_g = 1.65$ V (blue curve) a peak in the magnitude of the current emerges close to $V_b = 0$. These peaks occur when there is an alignment in energy of the localised state, $E_1$, with LLs in the two graphene layers[21]. In effect, the localised state in the hBN barrier acts as a "stepping stone" for electrons tunnelling between the Landau levels of the two graphene electrodes, see inset of Fig. 1c. Using our electrostatic model, we estimate that $\mu_B \approx 106$ meV and $\mu_T \approx 75$ meV when $V_b = 0$ and $V_g = 1.65$ V, which correspond to electron densities of $n_B \approx 7.26 \times 10^{11}$ cm$^{-2}$ and $n_T \approx 3.63 \times 10^{11}$ cm$^{-2}$ in the $Gr_B$ and $Gr_T$ layers. Thus, the resonant peak in the blue curve in Fig. 1c near $V_b = 0$ arises from resonant tunnelling through the defect's energy level $E_1$ and between the partially-filled $N_B = 2$ Landau level in $Gr_B$ and $N_T = 1$ in $Gr_T$, as shown schematically in the inset of Fig. 1c.

Figure 2a, b are colour plots of the differential conductance $G(V_g, V_b) = dI/dV_b$ at $T = 4.2$ K and $B = 0$ and 3.75 T. When $B = 0$, the intersecting pink/red curves corresponding to maxima in $G$ (see Fig. 2a) are the loci of the resonant tunnelling threshold

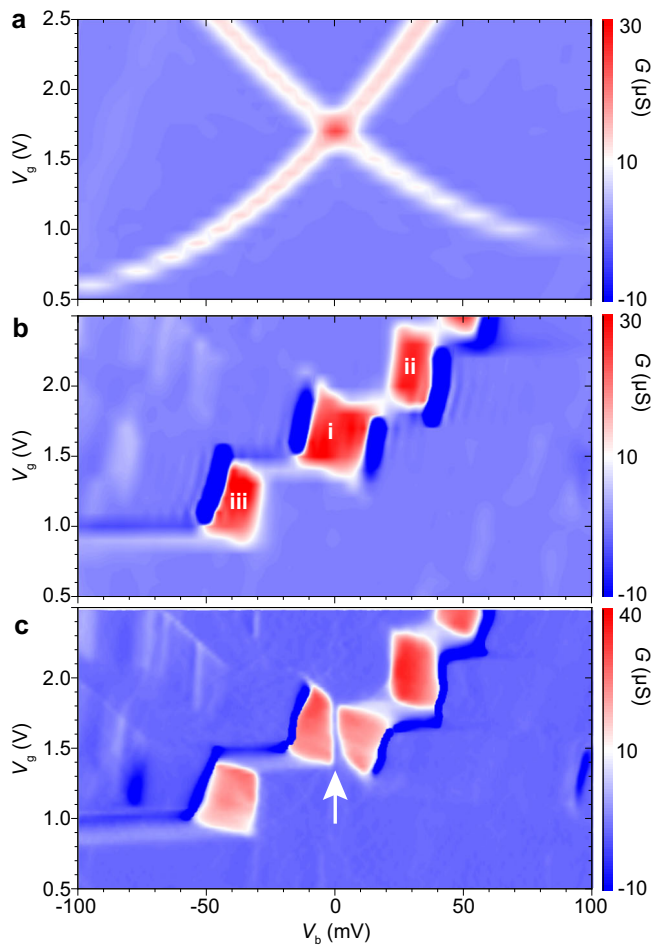

**Fig. 2 Emergence of a Coulomb gap in an applied magnetic field and at low temperatures.** Colour maps showing the dependence of the differential conductance, $G = dI/dV_b$ on applied bias voltage, $V_b$, and gate voltage $V_g$ when: **a** the magnetic field $B = 0$ T, and temperature $T = 4.2$ K; **b** $B = 3.75$ T, $T = 4.2$ K; and **c** $B = 3.75$ T, $T = 0.3$ K, with the formation of a Coulomb gap highlighted by the vertical white arrow. Colour bars show the amplitude of the conductance, $G$, in the colour map of each panel.

through the defect level $E_1$ at positive and negative bias voltages, corresponding to the step-like features in Fig. 1b[20]. When $B = 3.75$ T, these loci become a series of conductance maxima, shown in Fig. 2b where each maximum in conductance (red regions: (i), (ii), (iii)) is associated with a gradual change of the chemical potential $\mu_B$ while it is pinned within a LL. At this point, the LLs in the two electrodes are aligned with the energy level of the defect state and current can flow through the barrier. The blue gaps between these maxima correspond to the shift of $\mu_B$ into the neighbouring LLs of $Gr_B$ which requires a large change in $V_b$. In all three regions, electrons tunnel between the $N_B = 2$ LL in $Gr_B$ and $N_T = 1$ LL in $Gr_T$. The schematics of the electrostatic configuration of the three regions are shown in Supplementary Fig. 2.

Figure 2c shows the $G(V_b, V_g)$ map when $B = 3.75$ T at a much lower temperature, $T = 0.3$ K. We then observe a sharp suppression of $G(V_b)$ over a narrow voltage range ($\approx 5$ mV) centred on $V_b = 0$. This gap appears as a blue vertical stripe in the centre of region (i), indicated with a white arrow. This suppression of the conductance cannot be explained by splitting of the LLs caused by the lifting of the spin and valley degeneracy[12,21,22], as explained in Supplementary Note 2. Instead, we attribute the suppression to an intra-layer many-body interaction between the Fermi sea of electrons within the graphene collector layer and the injected electron, which tunnels out of the defect state. This type of interaction has been previously proposed and studied theoretically[30,31,34–36] to explain measurements of a similar magnetic-field induced Coulomb gap in the tunnel current between parallel two-dimensional electron gases (2DEGs) formed in GaAs quantum wells[28,29,32,33,37,38]. In a strong magnetic field, the electrons in the graphene layers become localised over a length scale determined by the radius of the cyclotron orbits and therefore electrons in the Fermi sea can become correlated at low temperatures. In a very strong magnetic field, we note that phases reminiscent to Wigner crystallisation can occur[40]. When the electron is injected into the graphene layer it creates a localised disruption within the Fermi sea which increases the energy of the system due the Coulomb interactions. The perturbation of the Fermi sea then spreads by the emission of collective excitations so that the system can relax to its final state. Therefore, in order for the electron to be injected into the Fermi sea it must have an energy greater than that which is lost during the relaxation process. In effect, the tunnelling electron must gain an energy, $U_c$, from the applied bias voltage which exceeds the difference between the energy of the disrupted (uncorrelated) Fermi sea and the final (correlated) relaxed state. The reciprocal process occurs for an electron leaving the graphene layer which leaves behind a localised hole and disruption of the Fermi sea. This many-body interaction results in a suppression of the tunnel current for bias voltages $|V_b| \lesssim U_c/e$. We discuss this mechanism and provide an estimate of $U_c$ and its dependence on magnetic field in the final section of the paper.

The blue curve in Fig. 3a shows in more detail the strong suppression of the conductance at $T = 0.3$ K and $V_g = 1.65$ V. The suppression of $G$ between the two maxima in conductance has a Coulomb gap width $\Delta V_b = 10 \pm 1$ mV, where the depth of the Coulomb gap is 25 μS. The red curve in Fig. 3a indicates clearly that the conductance peak around $V_b = 0$ is fully restored at 5 K. Furthermore, as indicated in the inset of Fig. 3a, the suppression of conductance is absent in zero magnetic field, even at 0.3 K. This indicates that the phenomenon has a "magnetic" origin.

Figure 3b is an Arrhenius plot of the temperature-dependent differential conductance $G(V_b = 0)$ at zero bias, $V_g = 1.65$ V and $B = 3.75$ T. The exponential growth of $G(V_b = 0)$ with increasing temperature from 1 K, indicates that the Coulomb gap closes due to temperature broadening for temperatures between ~1 K and 5 K. The activation energy, $E_a = k_B T_a$, where $T_a = 2$ K, quantifies the formation of low energy collective excitations in the Fermi liquid, as discussed later in the article. $G(V_b = 0)$ reaches a maximum at around 6 K and decreases slowly at higher temperatures. For $T > 9$ K the differential conductance decreases due to thermal broadening of the electron distribution around the Fermi energy.

**Magnetic field dependence of the Coulomb gap**. Figure 4a shows in more detail the $B$ field-dependence of the tunnel current $I(B)$ using a small but finite bias $V_b = 0.6$ mV, $V_g = 1.65$ V and at three different temperatures, 0.3, 2 and 7 K. The 3 peaks labelled $p = 1$, 2 and 3 in Fig. 4a occur at magnetic fields $B_p$, $B_1 = 3.85 \pm 0.02$ T, $B_2 = 1.97 \pm 0.02$ T and $B_3 = 1.30 \pm 0.02$ T. These magneto-oscillations are periodic in $1/B$ with a frequency $1/B_F = 1/(3.9 \pm 0.05)$ T$^{-1}$, see inset, where $B_F = pB_p$ and $B_p$ is the magnetic field value of the peak with index $p$. They are superficially similar in form to Shubnikov-de-Haas magneto-oscillations[24] but have a different physical origin. The sharp resonant peaks in the tunnel current occur at zero bias when the Landau levels in each graphene layer are aligned energetically with each other and with the defect energy level in the hBN layer, $E_1$, which is measured with respect to the Dirac point of the bottom graphene layer:

$$E_1 = E_{N_B}, \quad (1)$$

$$E_1 = E_{N_T} + \Delta E_D, \quad (2)$$

where $\Delta E_D = ebF_b$. The measurements in Fig. 4 were obtained for $V_g = 1.65$ V from which we estimate $\Delta E_D \approx 30$ meV and $E_1 \approx 100$ meV using the electrostatic model. To understand the origin of the $1/B$-periodic oscillations, we rewrite Eqs. (1) and (2) in terms of the magnetic field using the relation $E_N = \sqrt{2N}\hbar v_F v_F/l_B$, hence:

$$B = \frac{E_1^2}{2N_B e\hbar v_F^2}, \quad (3)$$

$$B = \frac{(E_1 - \Delta E_D)^2}{2N_T e\hbar v_F^2}. \quad (4)$$

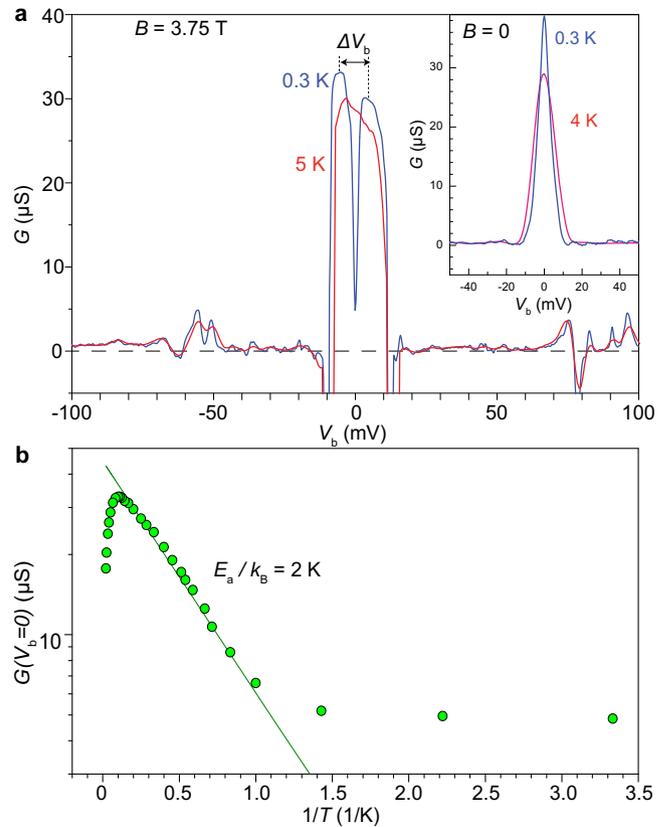

**Fig. 3 Temperature dependence of the conductance suppression in the Coulomb gap. a** Dependence of the conductance, $G$, on bias voltage, $V_b$, $G(V_b)$, when the applied gate voltage $V_g = 1.65$ V and magnetic field $B = 3.75$ T for temperatures of $T = 0.3$ K (blue) and $T = 5$ K (red). The width, $\Delta V_b$ of the dip in $G$ is measured as the distance between the peaks in $G$. Inset shows $G(V_b)$ when $V_g = 1.65$ V and $B = 0$ T for temperatures of $T = 0.3$ K (blue) and $T = 4$ K (red). **b** Arrhenius plot of $G(V_b = 0)$ as a function $1/T$ when $B = 3.75$ T showing the thermal activation behaviour with characteristic energy, $E_a$.

Combining Eqs. (3) and (4), we obtain the following ratio between the LL indices in the top and bottom layers, $N_T$ and $N_B$,

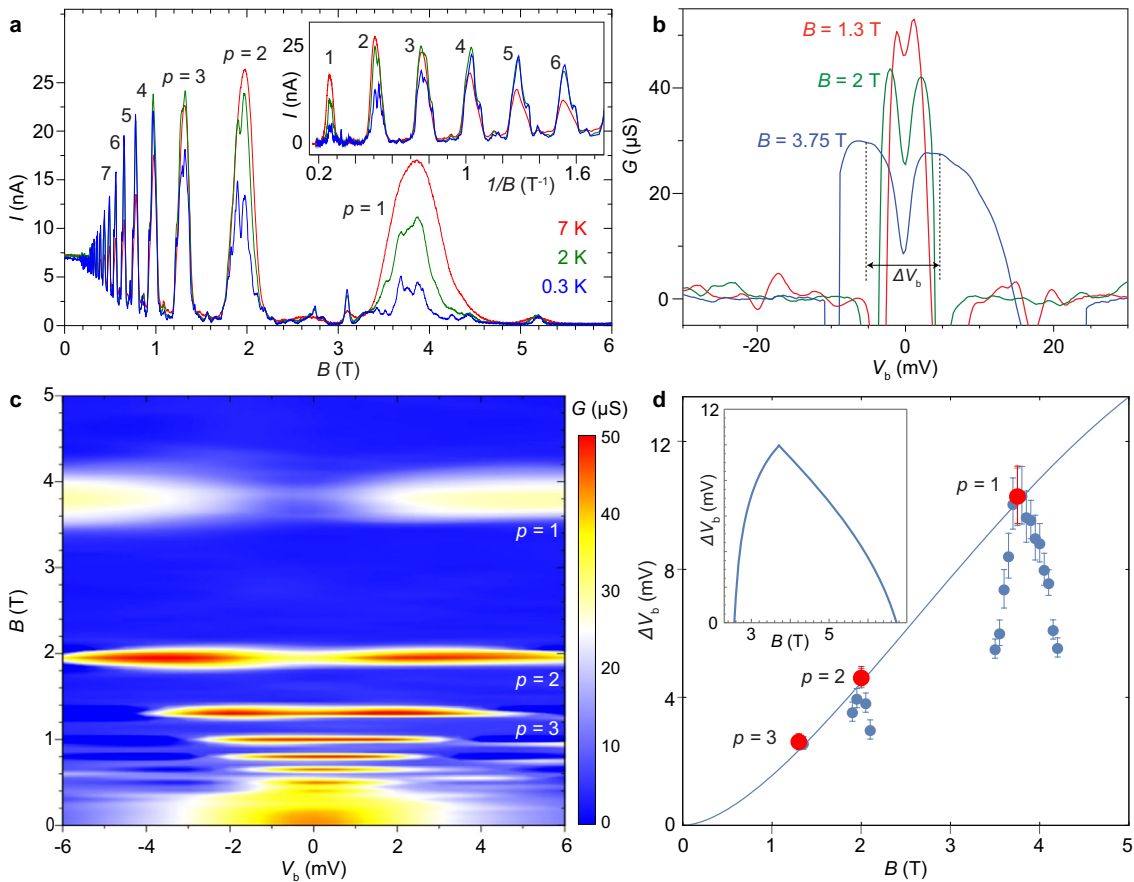

**Fig. 4 Magnetic field dependence of the Coulomb gap. a** The dependence of current, $I$, on magnetic field, $B$, $I(B)$ (and $I(1/B)$ inset), with applied gate, $V_g = 1.65$ V, and bias voltages $V_b = 0.6$ mV at temperatures $T = 7$ K (red), 2 K (green) and 0.3 K (blue). The peaks are labelled $p = 1, 2, 3$. **b** The conductance measured as a function of $V_b$, $G(V_b)$, when $V_g = 1.65$ V and $T = 0.3$ K for $B = 1.3, 2$ and 3.75 T. **c** Colour map of $G(V_b, B)$ measured when $V_g = 1.65$ V and $T = 0.3$ K. Colour bar shows the amplitude of the conductance in the colour map. **d** Red data points show the maximum in $\Delta V_b$ at $T = 0.3$ K for the $B$-fields shown in (**b**), the blue data points show the measured dependence of the gap width $\Delta V_b$ over the full range of $B$. The blue curve is a theoretical fit to the measurements given by Eq. (7). The inset shows the dependence $\Delta V_b(B)$ for the fractional filling of the Landau level with index 1 given by Eq. (9). Error bars estimate the uncertainty in determining the peak-to-peak width of the Coulomb gap from the measured data.

when their energies are aligned with the defect energy level:

$$\frac{N_T}{N_B} = \frac{n_T}{n_B} = \left(1 - \frac{\Delta E_D}{E_1}\right)^2. \quad (5)$$

On resonance, $\Delta E_D \approx 30$ meV and $E_1 \approx 100$ meV, and therefore $N_T/N_B = n_T/n_B \approx 1/2$. Consequently, as $B$ decreases the energy, $E_1$, of the defect state in the barrier becomes sequentially aligned with LLs $N_T = 1, 2, 3, 4...$ in the top layer and LLs with $N_B = 2, 4, 6, 8...$ in the bottom layer. The resonant condition described by Eqs. (1) and (2) can then be rewritten in the following form:

$$B_p = \frac{\Delta E_D^2}{2\hbar v_F^2 ep} \frac{1}{\left(\sqrt{2} - 1\right)^2}, \quad (6)$$

where $p = 1, 2, 3...$ labels the resonances corresponding to transitions between the $N_T = p$ and $N_B = 2p$ LLs. Equation (6) therefore corresponds to a series of peaks in $I(B)$ which are periodic in $1/B$ with frequency $1/B_F = 1/B_p(p = 1)$. It gives a value of $B_F = 4.0$ T which is in good agreement with our measured $B_F = 3.90 \pm 0.05$ T determined from the peak positions in Fig. 4a. We emphasise that the sequence of aligned LLs implied by Eq. (5) and which results in Eq. (6), arises due to the particular electrostatic configuration of our device. The series of resonant peaks would be modified for devices with a different structure and resonant condition.

The temperature dependence of the amplitudes of the magneto-oscillations are consistent with the emergence of a Coulomb gap. Their amplitudes at low fields, $B < 0.8$ T ($p > 3$), decrease with increasing temperature from 0.3 K to 7 K due to the temperature broadening of the density of states, see Fig. 4a and inset. In contrast, the amplitudes of the maxima with $p \leq 3$ decrease markedly with decreasing temperature due to the suppression of current around $V_b = 0$ by the formation of the Coulomb gap.

Figure 4b, c shows the measured line-shape of the $p = 1, 2$ and 3 resonances and the evolution of the Coulomb gap as a function of $V_b$ and $B$. We attribute the linewidths to the lifetime broadening of the LL states and the defect state. The dependence of the Coulomb gap width, $\Delta V_b$, on $B$ is indicated by the blue and red circles in Fig. 4d. The gap width is at a maximum when $B = B_p$ and the LLs are half-filled. The maximum value of $\Delta V_b$ for the $p = 1, 2$ and 3 resonance conditions and the corresponding value of $B_p$ are shown by the red circles in Fig. 4d. Our data demonstrate that the maximum value of $\Delta V_b$ decreases significantly with decreasing $B_p$.

**Analysis of the Coulomb gap dependence on $B$.** Our measured dependence of $\Delta V_b(B_p)$ can be compared with a semiclassical, hydrodynamic model[30] and complementary full quantum model[31], both of which describe the general case of injection of an electron into a 2DEG. The models yield the energy, $U_c$, which arises due to the interaction of an injected electron with a two-dimensional electron liquid and the subsequent relaxation of the liquid to the equilibrium state by the formation of collective excitations. In order to tunnel between the electrodes, an electron requires an energy, $U_c$, higher than the Fermi energy in the collector. The spectral density of this interacting system takes the form $A(E) \propto 1/\Gamma \exp(-(|E| - U_c)^2/4\Gamma^2)$, where $\Gamma$ is the energy broadening of the peaks[30]. This energy broadening depends on inhomogeneities of the electron gas caused by defects and impurities and also on the formation of low energy collective excitations due to thermal fluctuations. According to the semi-classical model[30], for half-filled LLs the peaks in $A(E)$ occur at $\pm U_c$ with:

$$U_c = \frac{\hbar v_F}{2\nu l_B^2 k_F} \ln(\nu r_s) = \frac{\hbar v_F}{4R_c^2 k_F} \ln(2k_F R_c r_s). \quad (7)$$

For graphene, the dimensionless parameter, $r_s = e^2/4\pi\varepsilon_0 \varepsilon \hbar v_F \approx 0.8$ with the background dielectric constant $\varepsilon = 2.8$ for graphene encapsulated by hBN[41,42]. The filling factor, $\nu = nh/eB$ and $R_c = \hbar k_F/eB$ is the classical cyclotron radius at the Fermi energy. For the measurements shown in Figs. 1–4, when the chemical potentials of the two graphene layers are aligned with the energy of the defect state at $V_b = 0$, the electron densities in the two graphene electrodes are $n_B \approx 7.26 \times 10^{11}$ cm$^{-2}$ and $n_T \approx 3.63 \times 10^{11}$ cm$^{-2}$. According to Eq. (7), $U_c$ is larger in the top graphene layer compared to its value in the bottom graphene layer. Therefore, we discount the influence of the Coulomb gap on the bottom layer when analysing the tunnel current. Consequently, on resonance and for small $V_b$, we assume the tunnel current takes the form:

$$I(V_b) \propto \int_0^{eV_b} A_{imp}(E - eV_b/2) A(E - eV_b) \Delta f(E, T, V_b) dE, \quad (8)$$

where $\Delta f(E, T, V_b) = f(E, T) - f(E - eV_b, T)$ and $f(E, T)$ is the Fermi-Dirac distribution function[33]. Here, $A_{imp}(E) \propto 1/\Gamma_{imp} \exp(-E^2/2\Gamma_{imp}^2)$ is the impurity spectral density with energy broadening $\Gamma_{imp}$. $I(V_b)$ has a maximum at $V_b = 2U_c/e$ but the position of the maximum in $G = dI/dV$ depends on the relative amplitudes of $\Gamma$ and $\Gamma_{imp}$. We make the assumption that these peaks in $G$ occur when $V_b = \pm U_c/e$ giving our measured value, $\Delta V_b \equiv 2U_c/e$[32]. The blue curve in Fig. 4d shows a plot of $2U_c/e$ calculated using Eq. (7) with $n = n_T$. The curve is in quantitative agreement with the maximum measured values of $\Delta V_b$, indicated by the red data points in Fig. 4d, in particular note the almost linear form of the blue curve between $B = 1$ and 4 T.

Our device enables us to measure the dependence of $\Delta V_b$ on the partial filling fraction, $\nu_f = \nu - 4N + 2$, of the $N_B = 1$ LL over a limited range of $\nu_f$, see blue filled circles in Fig. 4d. The finite energy broadening of the defect state and the density of states in the graphene electrodes means that we observe a current and a clear gap width $\Delta V_b$ even when $B \neq B_p$, in particular when $B$ is large. We find that $\Delta V_b$ is at a maximum when $B = B_p$ and $\nu_f = 2$, i.e. when the Landau level is half-filled. If the Landau level is only partially filled, then the separation between the electrons or holes in a given LL is much larger than $R_c$ and the interaction between the injected electron and the Fermi liquid is suppressed. As the LL is filled, there is stronger Coulomb repulsion between the injected electron and the Fermi liquid until this reaches a maximum at $\nu_f = 2$. Our results are consistent with the calculated dependence of $U_c$ on $\nu_f$, which was derived previously for two-dimensional electrons in a GaAs quantum well[31]. By modifying their formula for the case of graphene, we obtain the following dependence of $U_c$ on $\nu_f$:

$$U_c(\nu_f) = \frac{\hbar k_F \nu_F}{4N\nu} \ln\left((2N)^{3/2} r_s \sqrt{\nu_f}\right). \quad (9)$$

The inset of Fig. 4d shows the dependence of $\Delta V_b = 2U_c/e$ on $B$ with $U_c$ calculated using Eq. (9) when $p = 1$ (and $N = 1$). We find that the dependence provided by Eq. (9) is qualitatively consistent with the measured dependence of $\Delta V_b(\nu_f)$ given by the blue data points in Fig. 4d when $p = 1$. In particular, both the data and our model reveal the reduction of the Coulomb gap energy as the electron/hole liquid becomes more dilute. This is due to the weaker Coulomb interaction between the injected electron and the electron liquid. Moreover, the calculated line-shape in the inset of Fig 4d is in qualitative agreement with the measured data (blue data points in Fig. 4d). We emphasise that the quantum theoretical model in ref. [31] was not derived explicitly for graphene, nor for tunnelling through a defect state, yet it provides a physical insight for our data.

Previous studies of electrons tunnelling between the LLs of a double quantum well (DQW) GaAs/(AlGa)As heterostructure grown by molecular beam epitaxy, have also revealed the formation of a Coulomb gap[28,29,32,33,37,38]. In those devices, tunnelling occurs coherently between two parallel 2DEGs separated by a ~10 nm thick tunnel barrier, without an intermediate localised state. In ref. [32] over a range of filling factors for two different DQW devices, the Coulomb gap energy was shown to depend linearly on $B$, as we also observe over our range of $B$, see Fig. 4. In the DQW devices, it has been suggested that both inter- and intra-layer e-e interactions can affect the Coulomb gap[29,33,43,44]. In GaAs heterostructures the ratio $l_B/d \sim 1$ when $B = 1$ T, so that it is difficult to differentiate between the inter- and intra-layer origin of the Coulomb gap for coherent tunnelling in those devices. In our device, the twist angle misorientation of the two graphene layers means that the coherence between the initial and final states is suppressed. The strongly localised defect state induces the large k-vector shift required by the tunnelling electron. This type of device thus enables us to study the intra-layer e-e interaction as the effects of inter-layer interaction should be small.

We note that in our device the activation energy, $E_a \approx 0.2$ meV, is around an order of magnitude smaller than the width of the Coulomb gap, $e\Delta V_b$. A similar difference was also measured in DQW GaAs-(AlGa)As heterostructures[28,32,33,37,38] and has been investigated theoretically[30,34–36]. It can be explained by noting that as the temperature is increased, low energy collective excitations are formed in the electron liquid, for example phonons[34]. These excitations can assist tunnelling and the injected electron can tunnel into the resulting inhomogenous electron liquid with a lower energy. This gives rise to the temperature-induced broadening of the features in the spectral density at $\pm U_c$. The measured thermal activation energy $E_a$ therefore corresponds to the formation energy of the low energy collective excitations in the system rather than the Coulomb gap energy[30,33–36].

## Conclusion

In summary, we have investigated defect-assisted resonant tunnelling using intentionally misaligned graphene/hBN/graphene tunnel transistors. The tunnel current flowing through a single strongly localised defect state in the hBN barrier provides a spectroscopic probe of the Landau level density of states in the graphene layers. A strong suppression of the tunnel conductance is observed in the presence of a quantising magnetic field applied perpendicular to the layers. Analysis and modelling of the measured temperature and magnetic field dependences of the tunnel current indicate that the observed suppression of the differential conductance arises from e-e interactions which form a magnetically-induced Coulomb gap in the tunnelling density of states of the graphene layers at low temperatures. This suppression is in marked contrast to the previously reported strong increase in tunnel current due to a Fermi edge singularity in which e-e interactions enhance the tunnel current flowing between a two-dimensional electron gas and a strongly localised state in zero magnetic field[45]. Our analysis accounts for the magnetic field dependence of the width of the Coulomb gap in graphene. The presence of the localised state in the barrier, which acts as an electron "injector", provides an innovative technique to explore this effect and offers insights into these many-body quantum phenomena in graphene-based devices.

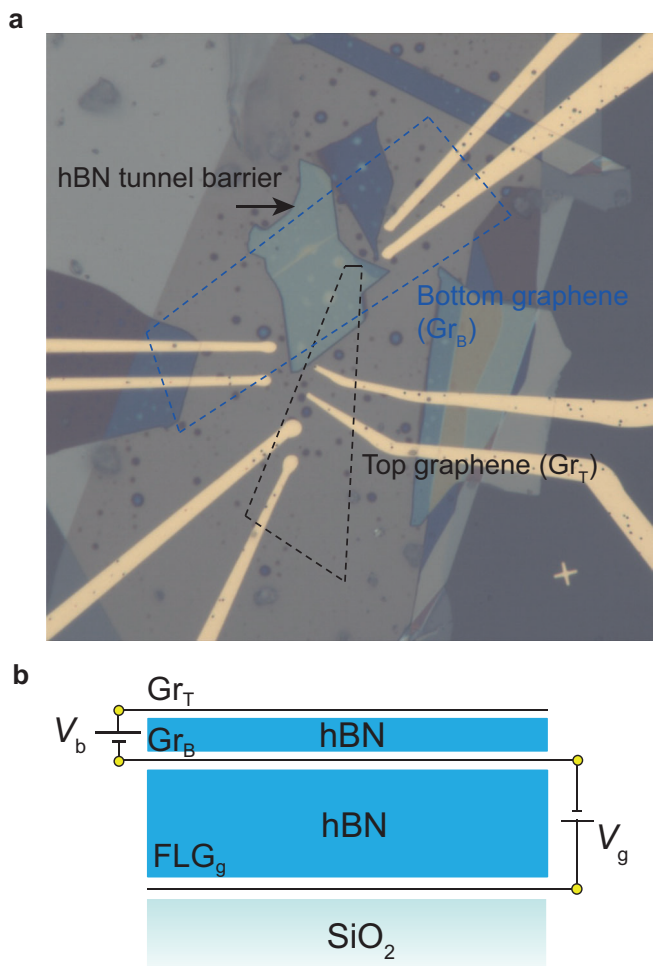

**Fig. 5 Configuration of the tunnelling device. a** Optical microscopy image of our device. The positions of the bottom and top graphene monolayers are indicated with blue and black dashed lines respectively. The intersection area of the graphene layers (area of the tunnelling device) is ~50 μm², **b** is a schematic diagram showing the top $Gr_T$ and bottom $Gr_B$ graphene layers and the configuration of the applied gate, $V_g$, and bias $V_b$ voltages. The few-layer graphene gate (FLG) is located between the $SiO_2$ and thick hBN layer.

## Methods

Figure 5a shows an optical microscopy image of the device studied in this article. The van der Waals heterostructure was fabricated using the dry transfer technique[9]. The contacts to graphene layers were fabricated using electron beam lithography and Cr/Au deposition. The crystal lattices of graphene layers were rotated relative to each other at an angle of about 30°. $Gr_T$ and $Gr_B$ in the figure denote the upper and lower graphene monolayers separated by a hBN barrier. In Fig. 5b we show a schematic diagram of the layer and voltage configuration of our heterostructure. A 30 nm-thick graphite (few layer graphene) layer acts as the gate electrode which insulated from the rest of the structure using a 27 nm thick hBN insulating layer. The whole structure is placed on a 0.5 mm-thick quartz substrate. The current flow between the $Gr_B$ and $Gr_T$ was measured using a 4-probe configuration with Cr/Au electrodes on each of the $Gr_B$ and $Gr_T$ layers, see Fig. 5a.

### Data availability
Source data are provided at https://doi.org/10.17028/rd.lboro.23301866. Other data that support the plots within this paper and other findings of this study are available from the corresponding authors upon reasonable request.

### References
1. Wallbank, J. R. et al. Excess resistivity in graphene superlattices caused by umklapp electron-electron scattering. *Nat. Phys.* **15**, 32–36 (2019).
2. Bandurin, D. A. et al. Negative local resistance caused by viscous electron backflow in graphene. *Science* **351**, 1055 (2016).
3. Cao, Y. et al. Unconventional superconductivity in magic-angle graphene superlattices. *Nature* **556**, 43–50 (2018).
4. Cao, Y. et al. Correlated insulator behaviour at half-filling in magic-angle graphene superlattices. *Nature* **556**, 80–84 (2018).
5. Gorbachev, R. V. et al. Strong Coulomb drag and broken symmetry in double-layer graphene. *Nat. Phys.* **8**, 896 (2012).
6. Liu, X., Watanabe, K., Taniguchi, T., Halperin, B. I. & Kim, P. Quantum Hall drag of exciton condensate in graphene. *Nat. Phys.* **13**, 746 (2017).
7. Britnell, L. et al. Field-effect tunneling transistor based on vertical graphene heterostructures. *Science* **335**, 947–950 (2012).
8. Feenstra, R. M., Jena, D. & Gu, G. Single-particle tunneling in doped graphene-insulator-graphene junctions. *J. Appl. Phys.* **111**, 043711 (2012).
9. Mishchenko, A. et al. Twist-controlled resonant tunnelling in graphene/boron nitride/graphene heterostructures. *Nat. Nanotechnol.* **9**, 808–813 (2014).
10. Greenaway, M. T. et al. Resonant tunnelling between the chiral Landau states of twisted graphene lattices. *Nat. Phys.* **11**, 1057–1062 (2015).
11. Wallbank, J. R. et al. Tuning the valley and chiral quantum state of Dirac electrons in van der Waals heterostructures. *Science* **353**, 575–579 (2016).
12. Khanin, Yu. N. et al. Observation of spin and valley splitting of Landau levels under magnetic tunneling in graphene/boron nitride/graphene structures. *JETP Lett.* **107**, 238 (2018).
13. Kuzmina, A. et al. Resonant light emission from graphene/hexagonal boron nitride/graphene tunnel junctions. *Nano Lett.* **21**, 8332–8339 (2021).
14. Lin, K. et al. Emergence of interlayer coherence in twist-controlled graphene double layers. *Phys. Rev. Lett.* **129**, 187701 (2022).
15. Inbar, A et al. The quantum twisting microscope. *arXiv* https://doi.org/10.48550/arXiv.2208.05492 (2022).
16. Jung, S. et al. Vibrational properties of h-BN and h-BN-graphene heterostructures probed by inelastic electron tunneling spectroscopy. *Sci. Rep.* **5**, 16642 (2015).
17. Vdovin, E. E. et al. Phonon-assisted resonant tunneling of electrons in graphene-boron nitride transistors. *Phys. Rev. Lett.* **116**, 186603 (2016).
18. Chandni, U., Watanabe, K., Taniguchi, T. & Eisenstein, J. P. Signatures of phonon and defect-assisted tunneling in planar metal-hexagonal boron nitride-graphene junctions. *Nano Lett.* **16**, 7982–7987 (2016).
19. Chandni, U., Watanabe, K., Taniguchi, T. & Eisenstein, J. P. Evidence for defect-mediated tunneling in hexagonal boron nitride-based junctions. *Nano Lett.* **15**, 7329–7333 (2015).
20. Greenaway, M. T. et al. Tunnel spectroscopy of localised electronic states in hexagonal boron nitride. *Commun. Phys.* **1**, 94 (2018).
21. Khanin, Yu. N. et al. Tunneling in graphene/h-BN/graphene heterostructures through zero-dimensional levels of defects in h-BN and their use as probes to measure the density of states of graphene. *JETP Lett.* **109**, 482 (2019).
22. Keren, I. et al. Quantum-dot assisted spectroscopy of degeneracy-lifted Landau levels in graphene. *Nat. Commun.* **11**, 1–9 (2020).
23. Papadopoulos, N. et al. Tunneling spectroscopy of localized states of $WS_2$ barriers in vertical van der Waals heterostructures. *Phys. Rev. B* **101**, 165303 (2020).
24. Zheng, S. et al. Robust quantum oscillation of Dirac fermions in a single-defect resonant transistor. *ACS Nano* **15**, 20013–20019 (2021).
25. Devidas, T. R., Keren, I. & Steinberg, H. Spectroscopy of $NbSe_2$ using energy-tunable defect-embedded quantum dots. *Nano Lett.* **21**, 6931–6937 (2021).
26. Seo, Y. et al. Defect-assisted tunneling spectroscopy of electronic band structure in twisted bilayer graphene/hexagonal boron nitride moiré superlattices. *Appl. Phys. Lett.* **120**, 203103 (2022).

27. Ashoori, R. C., Lebens, J. A., Bigelow, N. P. & Silsbee, R. H. Equilibrium tunneling from the two-dimensional electron gas in GaAs: evidence for a magnetic-field-induced energy gap. *Phys. Rev. Lett.* **64**, 681–684 (1990).
28. Eisenstein, J. P., Pfeiffer, L. N. & West, K. W. Coulomb barrier to tunneling between parallel two-dimensional electron systems. *Phys. Rev. Lett.* **69**, 3804 (1992).
29. Eisenstein, J. P., Pfeiffer, L. N. & West, K. W. Evidence for an interlayer exciton in tunneling between two-dimensional electron systems. *Phys. Rev. Lett.* **74**, 1419 (1995).
30. Aleiner, I. L., Baranger, H. U. & Glazman, L. I. Tunneling into a two-dimensional electron liquid in a weak magnetic field. *Phys. Rev. Lett.* **74**, 3435 (1995).
31. Aleiner, I. L. & Glazman, L. I. Two-dimensional electron liquid in a weak magnetic field. *Phys. Rev. B* **52**, 11296 (1995).
32. Turner, N. et al. Tunneling between parallel two-dimensional electron gases. *Phys. Rev. B* **54**, 10614 (1996).
33. Reker, T. et al. Magnetic-field-induced suppression of tunnelling into a two-dimensional electron system. *J. Phys. Condens. Matter* **14**, 5561 (2002).
34. Johansson, P. & Kinaret, J. M. Magnetophonon shakeup in a Wigner crystal: applications to tunneling spectroscopy in the quantum Hall regime. *Phys. Rev. Lett.* **71**, 1435 (1993).
35. Johansson, P. & Kinaret, J. M. Tunneling between two two-dimensional electron systems in a strong magnetic field. *Phys. Rev. B* **50**, 4671 (1994).
36. Haussmann, R. Electronic spectral function for a two-dimensional electron system in the fractional quantum Hall regime. *Phys. Rev. B* **53**, 7357 (1996).
37. Dolgopolov, V. T., Drexler, H., Hansen, W., Kotthaus, J. P. & Holland, M. Electron correlations and Coulomb gap in a two-dimensional electron gas in high magnetic fields. *Phys. Rev. B* **51**, 7958 (1995).
38. Deviatov, E. V., Shashkin, A. A., Dolgopolov, V. T., Hansen, W. & Holland, M. Tunneling measurements of the Coulomb pseudogap in a two-dimensional electron system in a quantizing magnetic field. *Phys. Rev. B* **61**, 2939 (2000).
39. Deshpande, M. R., Sleight, J. W., Reed, M. A., Wheeler, R. G. & Matyi, R. J. Spin splitting of single 0D impurity states in semiconductor heterostructure quantum wells. *Phys. Rev. Lett.* **76**, 1328 (1996).
40. Zhou, H., Polshyn, H., Taniguchi, T., Watanabe, K. & Young, A. F. Solids of quantum Hall skyrmions in graphene. *Nat. Phys.* **16**, 154–158 (2020).
41. Hwang, E. H. & Sarma, S. D. Dielectric function, screening, and plasmons in two-dimensional graphene. *Phys. Rev. B* **75**, 205418 (2007).
42. Kim, K. K. et al. Synthesis and characterization of hexagonal boron nitride film as a dielectric layer for graphene devices. *ACS Nano* **6**, 8583–8590 (2012).
43. Levitov, L. S. & Shytov, A. V. Spatial coherence of tunneling in double wells. *arXiv* https://doi.org/10.48550/arXiv.cond-mat/9507058 (1995).
44. Chowdhury, D., Skinner, B. & Lee, P. A. Semiclassical theory of the tunneling anomaly in partially spin-polarized compressible quantum Hall states. *Phys. Rev. B* **97**, 195114 (2018).
45. Geim, A. K. et al. Fermi-edge singularity in resonant tunneling. *Phys. Rev. Lett.* **72**, 2061 (1994).

## Acknowledgements

This work has been funded by the Engineering and Physical Sciences Research Council [Grant numbers EP/S030719/1, EP/V007033/1, and EP/V008110/1]. We acknowledge support from the Horizon 2020 EC-FET Core 3 European Graphene Flagship Project, EC-FET Quantum Flagship Project 2D-SIPC and Lloyd Register Foundation Nano-technology Grant. K.S.N. is grateful to the Ministry of Education, Singapore, for their support under its Research Centre of Excellence award to the Institute for Functional Intelligent Materials (I-FIM, project No. EDUNC-33-18-279-V12) and to the support from the Royal Society (UK, grant number RSRP\R\190000). The device was fabricated at the University of Manchester by Dr Y. Cao. E.E.V., Y.N.K. and S.V.M. were supported by the Russian Ministry of Science and Higher Education (Grant No. 075-01304-23-00).

## Author contributions

The measurements were carried out by E.E.V., Y.N.K., S.V.M., O.M., A.P., A.M. and L.E. Theoretical analysis was undertaken by M.T.G., S.S. and V.I.F. The data were analysed by E.E.V., M.T.G., Y.N.K., S.V.M., S.S., A.M., A.K.G., K.S.N. and L.E. The manuscript was prepared by E.E.V., M.T.G. and L.E. with input from all of the other authors.

## Competing interests

The authors declare no competing interests.

## Additional information

**Supplementary information** The online version contains supplementary material available at https://doi.org/10.1038/s42005-023-01277-y.

**Correspondence** and requests for materials should be addressed to Mark T. Greenaway.

**Peer review information** *Communications Physics* thanks Van-Nham Phan, Hadar Steinberg and the other, anonymous, reviewer(s) for their contribution to the peer review of this work.

**Reprints and permission information** is available at http://www.nature.com/reprints

**Publisher's note** Springer Nature remains neutral with regard to jurisdictional claims in published maps and institutional affiliations.

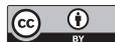